\documentclass[conference]{IEEEtran}
\IEEEoverridecommandlockouts
\usepackage{cite}
\usepackage{amsmath,amssymb,amsfonts}
\usepackage{algorithmic}
\usepackage{graphicx}
\usepackage{textcomp}
\usepackage{xcolor}
\usepackage{physics}
\usepackage{hyperref}
\usepackage[T1]{fontenc}
\usepackage[a4paper, total={6in, 8in}, left=14mm, right=14mm, top=20mm, bottom=46mm]{geometry}
\def\BibTeX{{\rm B\kern-.05em{\sc i\kern-.025em b}\kern-.08em
    T\kern-.1667em\lower.7ex\hbox{E}\kern-.125emX}}
\begin{document}

\title{Asynchronous Telegate and Teledata Protocols for Distributed Quantum Computing
\thanks{The first-named author was supported by a lectureship at Saskatchewan Polytechnic. The third-named author was partially supported by an Natural Sciences and Engineering Research Council of Canada (NSERC) Discovery Grant, a Canadian Tri-Agency New Frontiers in Research (NFRF) Exploration Stream Grant, and a Pacific Institute for the Mathematical Sciences (PIMS) Collaborative Research Group Award.}
}

\author{\IEEEauthorblockN{Jacob Peckham}
\IEEEauthorblockA{\textit{Dept. of Computer Science}\\
\textit{University of Saskatchewan}\\
Saskatoon, Canada \\
jep243@usask.ca}
\and
\IEEEauthorblockN{Dwight Makaroff}
\IEEEauthorblockA{\textit{Dept. of Computer Science}\\
\textit{University of Saskatchewan}\\
Saskatoon, Canada \\
makaroff@cs.usask.ca}
\and
\IEEEauthorblockN{Steven Rayan}
\IEEEauthorblockA{\textit{Centre for Quantum Topology and Its Applications (quanTA)}
\IEEEauthorblockA{\textit{Dept. of Mathematics and Statistics}}
\textit{University of Saskatchewan}\\
Saskatoon, Canada \\
rayan@math.usask.ca}
}

\maketitle

\begin{abstract}
The cost of distributed quantum operations such as the telegate and teledata protocols is high due to latencies from distributing entangled photons and classical information. This paper proposes an extension to the telegate and teledata protocols to allow for asynchronous classical communication which hides the cost of distributed quantum operations. We then discuss the benefits and limitations of these asynchronous protocols and propose a quantum network card as an example of how asynchronous quantum operations might be used.
\end{abstract}


\section{Introduction and Motivation}
Quantum computers promise the ability to solve problems that classical computers have no polynomial-time solution for, and they can solve them in polynomial time \cite{divincenzo1995compute}. While quantum computers promise to offer efficient solutions to classically difficult problems, current quantum computers in the Noisy Intermediate Scale Quantum (NISQ) era are currently not capable of solving many of these problems \cite{lao2021}. One proposed strategy to effectively scale quantum computing systems is to use distributed systems. With a distributed quantum computer, the number of qubits is expected to scale linearly with the number of nodes in the system which could result in exponential computational speed-ups for certain classes of problems \cite{cuomo2020}.\par

Distributed quantum computing requires a method for transmitting quantum data or operations across some form of network. Two common methods for accomplishing this are the telegate and teledata protocols \cite{caleffi2022}. Both the telegate and teledata protocols require the transmission of a single pair of entangled photons and two bits of classical information. Latency from distributing the entangled photons and classical information make distributed operations expensive. Lengthy transmission times can cause loss of information due to the short decoherence times in the transmitted photons and the qubits on the target device \cite{illiano2022}. This paper proposes asynchronous variations of the telegate and teledata protocols to reduce the time constraints on classical communication in distributed quantum computing. \par

This paper will be broken into several sections; first, some background will be given on distributed quantum computing and the quantum internet, as well the telegate and teledata protocols. Next, the asynchronous telegate and teledata protocols will be described as well as a discussion on their benefits and limitations. Finally, a Quantum Network Card (QNC) is proposed to show a potential network interface for asynchronous distributed quantum operations.\\

\emph{Note Regarding Current Article Version.} In the prior version of this article on the arXiv and in the version accepted for publication in the proceedings of "Unlocking Quantum Potentials: Advances in Computing and Communication Technologies" (a workshop of the 10\textsuperscript{th} IEEE World Forum on the Internet of Things [IEEE WFIoT2024]), we conveyed proposals for asynchronous telegate and teledata protocols that do not require the use of nonunitary operators. Since then, we have run additional experiments on simulators and on IBM Brisbane that reveal the need for such operators to ensure expected outputs. The prior versions of the asynchronous protocols have been revised accordingly. We will note that the nonunitary aspects of these protocols may necessitate new hardware and/or software paradigms; we do not speculate upon these here. \par

\section{Background}\label{BGSection}
Distributed quantum computing has a variety of applications such as: scaling quantum computers \cite{cuomo2020}, quantum key distribution \cite{sharma2021}, and quantum clock synchronization \cite{nande2024}. The DiVincenzo Criteria \cite{divincenzo2000criteria} lays out the general requirements needed for quantum computing, but it also states that for quantum communication, there are two primary requirements: 
\begin{itemize}
\item the ability to convert stationary qubits to flying qubits and flying qubits to stationary qubits; and
\item the ability to reliably transmit flying qubits between specified locations.
\end{itemize}
Stationary qubits are qubits which cannot be moved, such as superconducting qubits, and flying qubits are qubits which can be moved or transmitted. There have been several mediums used for flying qubits such as microwave photons, optical photons, and surface acoustic wave phonons \cite{zhong2021}. However, flying qubits are often implemented using optical photons to allow for their transmission on existing optical networks \cite{awschalom2021, herbauts2013, sharma2021}. \par

When entangled photons are transmitted down optical fiber, the successful rate of transmission drops off exponentially with the distance the photons are transmitted. To solve this issue, quantum repeaters are added to extend the range that entangled photons can be transmitted \cite{pant2019}. On top of the requirements described by DiVincenzo for the transmission of quantum states, quantum networks will also need to support the transmission of classical information \cite{dahlberg2019}. This classical communication is required for protocols like telegate, but it is also required for keeping track of important meta-data such as qubit ID, creation time, and decoherence time \cite{pompili2022}. This has led to a number of proposals on effective ways to use networks made of quantum repeaters for both end-to-end entanglement \cite{wang2022, van2013}, as well as multi-partite entangled states \cite{pirker2018, pirker2019, epping2017}. \par

While large-scale examples of quantum networks do not yet exist, there have been a number of small proof-of-concept networks that have been implemented, including: entanglement established over 1.3 km \cite{kozlowski2020}; entangled photons being routed over a small software-defined network \cite{herbauts2013}; and microwave photons being used to transfer quantum states between superconducting qubits \cite{magnard2020}. Paradigms have also been developed for designing architectures for distributed quantum computers \cite{meter2006, nickerson2014} and for general networking stacks for the eventual quantum internet \cite{dahlberg2019, van2022, kozlowski2020}. \par

Other work has been done on topics such as asynchronous entanglement distribution \cite{yang2024, wang2022}, where the focus is on the network stack and distributing entangled photons. Other protocols have been designed for distributed quantum operations seeking to improve the efficiency of entangled photon usage \cite{stahlke2011, wu2023}, and looking at the classical communication costs of distributed quantum operations \cite{parakh2022, lo2000}. To the best of our knowledge, there are currently no methods proposed for asynchronous telegate and teledata protocols, where the focus is on allowing for local operations to continue while the distributed operation finishes.

\section{Telegate and Teledata}\label{TTSection}
\begin{figure}[!t]
	\centering
	\includegraphics[scale=0.25]{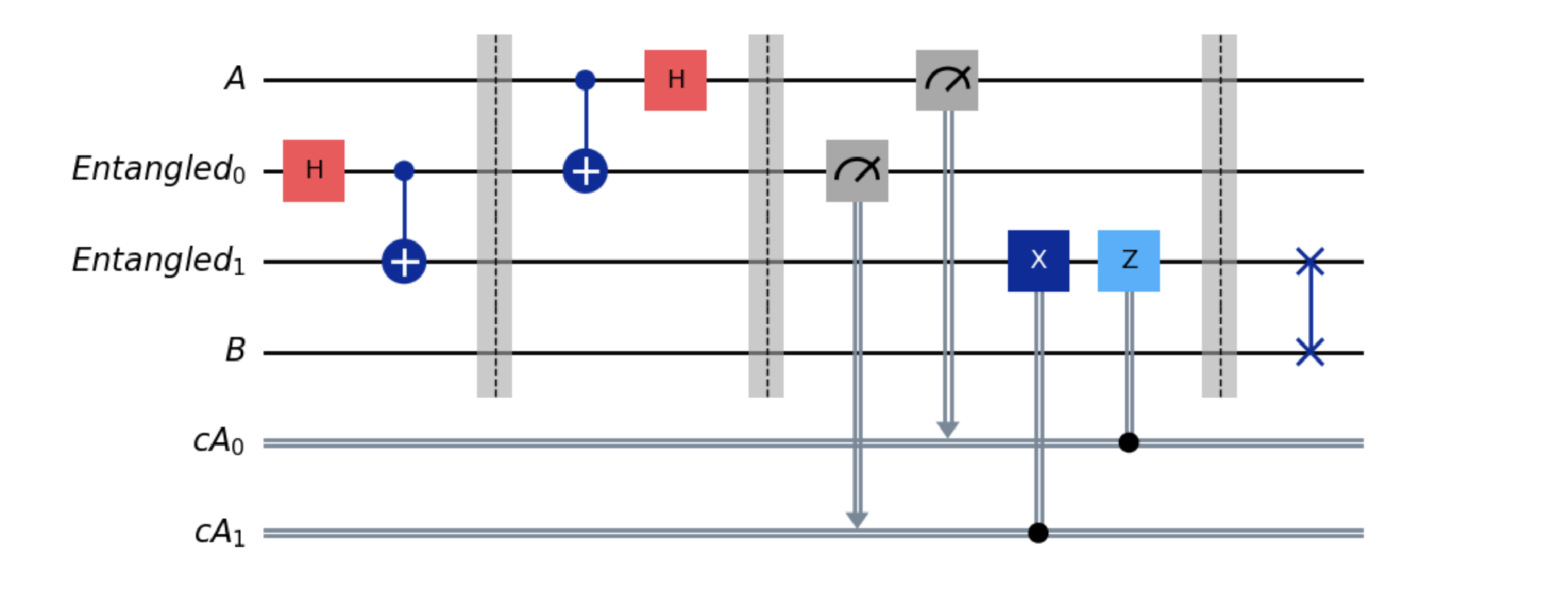}
	\includegraphics[scale=0.25]{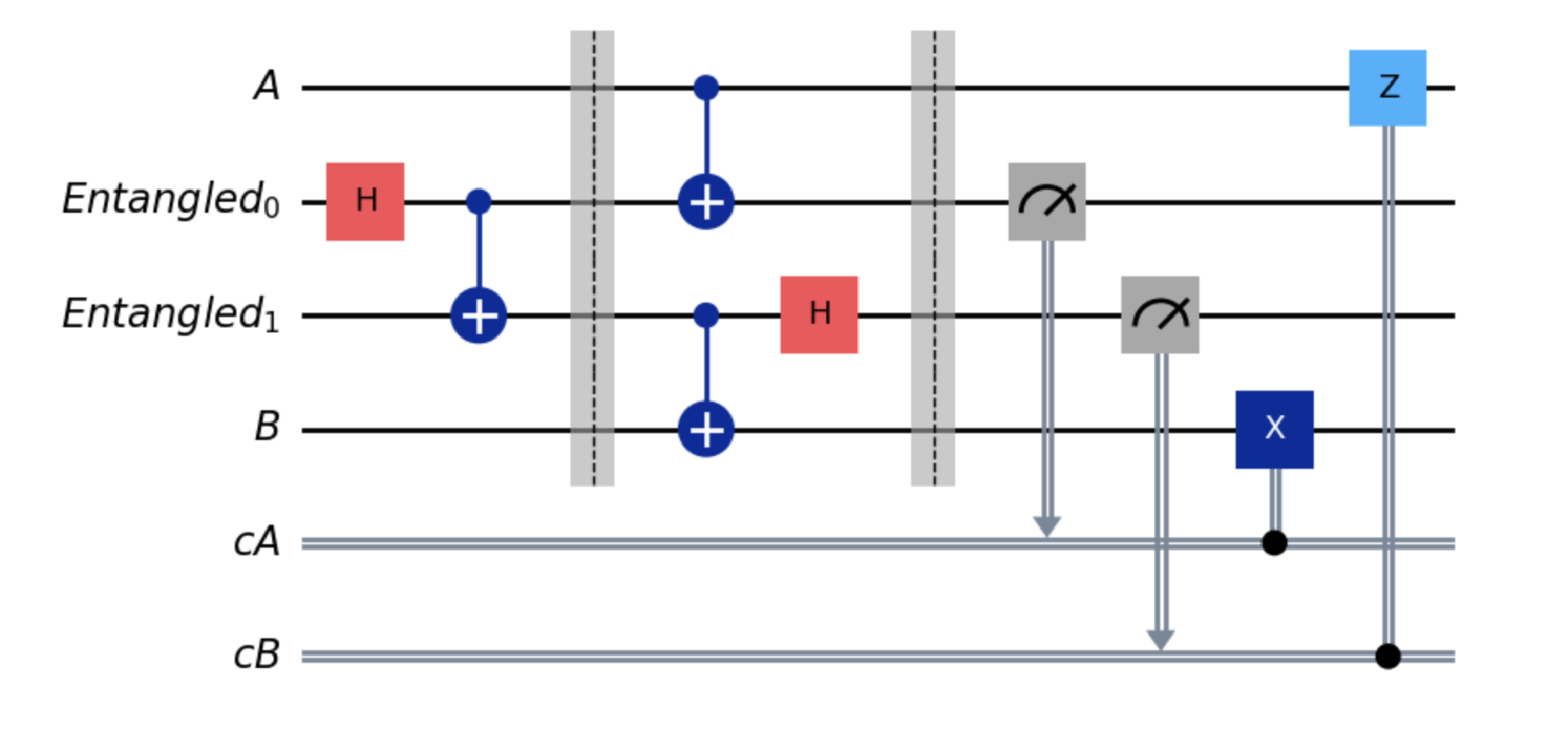}
	\caption{\textit{These are quantum circuits made with Qiskit showing teledata (top) and telegate (bottom) operations. The qubits Entangled represent an entangled state that is transmitted to two different quantum computers, with $Entangled_0$ going to computer A and $Entangled_1$ going to computer B. The classical bits cA start at computer A and classical bits cB start at computer B. After measurement, cA is transmitted to computer B, and cB is transmitted to computer A. Each circuit is divided into sections which are marked by the vertical lines.}}
	\label{fig:quantum_distributed_operations}
\end{figure}
The teledata and telegate protocols are based off of the principles of quantum teleportation, which uses a pair of maximally entangled photons and two bits of classical information to move a state from one quantum device to another. Both the teledata and telegate protocols can be seen in Figure \ref{fig:quantum_distributed_operations}. The telegate protocol will transmit the state in qubit $A$ to qubit $B$, and in the process qubit $A$ will no longer contain its original state. The telegate protocol performs a distributed CNot operation with qubit $A$ being the control and qubit $B$ being the target.\par

For both telegate and teledata, the first section of the circuit is to establish the entangled photons in the state $\ket{\Phi} = \frac{\ket{00} + \ket{11}}{\sqrt{2}}$; however, in an actual system the entangled photons would be generated outside the circuit and delivered to the appropriate quantum computing devices by the quantum network. After the entanglement is generated, the teledata and telegate protocols differ slightly. In the teledata protocol, the qubit $A$ is in some arbitrary state $\ket{\psi} = \alpha\ket{0} + \beta\ket{1}$, and the qubit $B$ is in the state $\ket{0}$. In the second section of the circuit, the state for qubit $A$ and the two entangled photons is:
\begin{equation}
\begin{aligned}
\ket{\Psi} = \frac{1}{2}(&\ket{00}_A(\alpha\ket{0} + \beta\ket{1}) + \\
			 &\ket{01}_A(\alpha\ket{1} + \beta\ket{0}) + \\
			 &\ket{10}_A(\alpha\ket{0} - \beta\ket{1}) + \\
			 &\ket{11}_A(\alpha\ket{1} - \beta\ket{0}))
\end{aligned}
\end{equation}
The qubits with the subscript $A$ correspond, respectively, with the first two qubits in the circuit that will be measured. By applying $X$ and $Z$ gates to the second entangled photon based on the measurement in the first two qubits, its state will be changed to $\alpha\ket{0} + \beta\ket{1}$, which is just the state $\ket{\psi}$ from qubit $A$. Then, after applying a CNot gate in the last step, $\ket{\psi}$ is transferred to the target qubit $B$. \par

The telegate protocol is a little different. Qubit $A$ is still in the arbitrary state $\ket{\psi}$, but qubit $B$ is also in a arbitrary state $\ket{\phi}$. Here, the resulting state before measurement is:
\begin{equation}
\begin{aligned}
\ket{\Psi} = \frac{1}{2}(&\ket{00}_e(\alpha\ket{0}\ket{\phi} + \beta\ket{1}\bar{\ket{\phi}}) + \\
			 &\ket{01}_e(\alpha\ket{0}\ket{\phi} - \beta\ket{1}\bar{\ket{\phi}}) + \\
			 &\ket{10}_e(\alpha\ket{0}\bar{\ket{\phi}} + \beta\ket{1}\ket{\phi}) + \\
			 &\ket{11}_e(\beta\ket{1}\ket{\phi} - \alpha\ket{0}\bar{\ket{\phi}}))
\end{aligned}
\end{equation}
The qubits with the subscript $e$ represent the entangled photons which will be measured, and $\bar{\ket{\phi}}$ is the resulting state from applying the $X$ gate to $\ket{\phi}$. After measurement and the conditional application of the $X$ and $Z$ gates, the state of the qubits $A$ and $B$ will be in the state $\alpha\ket{0}\ket{\phi} + \beta\ket{1}\bar{\ket{\phi}}$, which is the same as if only a CNot gate had been applied.

\section{Asynchronous Telegate and Teledata} \label{AsyncSection}
\begin{figure}
	\centering
	\includegraphics[scale=0.23]{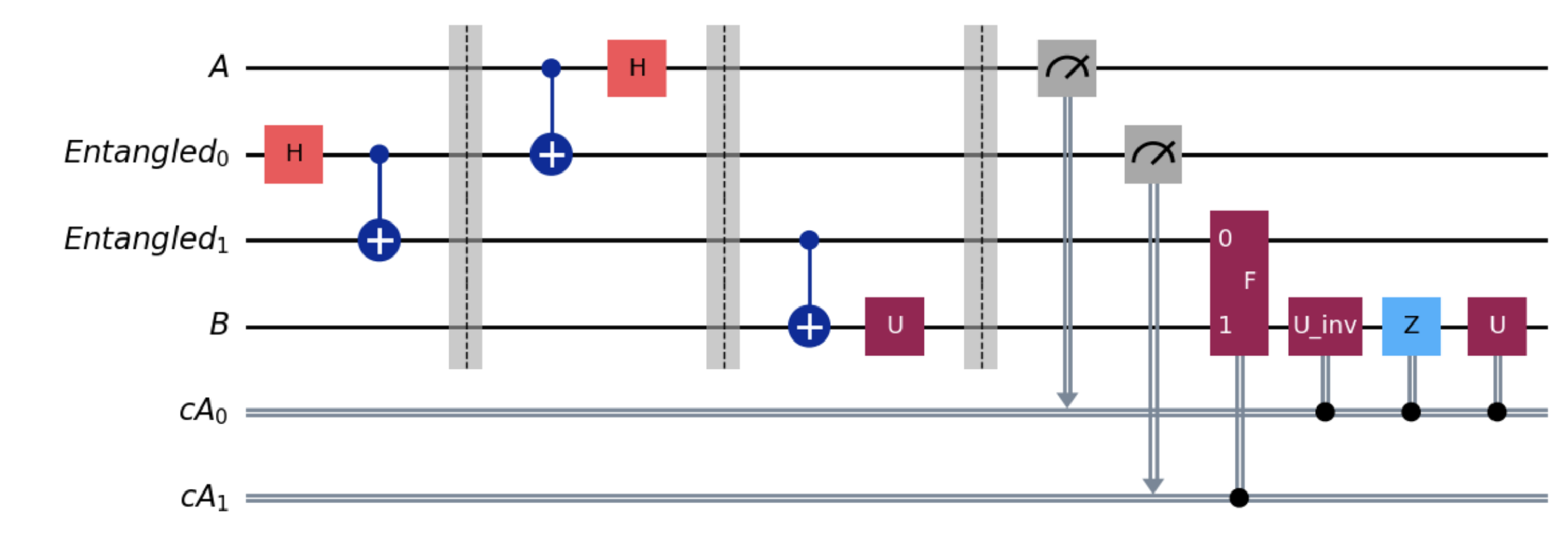}
	\includegraphics[scale=0.23]{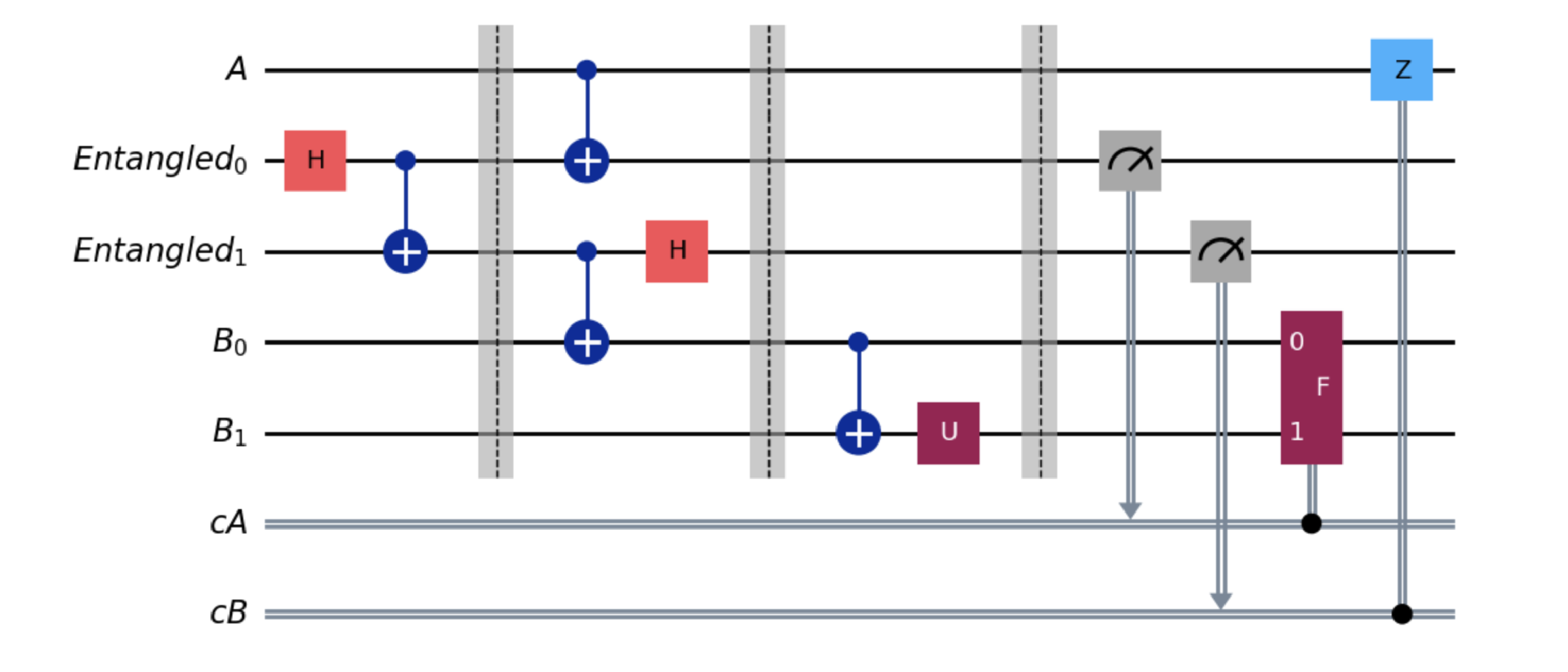}
	\caption{\textit{These are quantum circuits made with Qiskit showing async-teledata (top) and async-telegate (bottom) operations. The gates labelled ``U'' represent arbitrary unitary operations applied to the corresponding qubit.}}
	\label{fig:async_operations}
\end{figure}
The asynchronous versions of the teledata and telegate operations can be seen in Figure \ref{fig:async_operations}. The asynchronous versions of teledata and telegate share the same basic structure as the standard versions, but include an additional section before measurement containing a CNot operation and an arbitrary unitary operation. Since a composition $U=U_nU_{n-1}...U_1$ of unitary operators is again unitary, the validity of the asynchronous protocols can be shown by applying an arbitrary unitary operator in the circuit before measurement. Then, arbitrary local operations can take place before measurement and classical information transfer is completed. \par

In Section \ref{BGSection}, the asynchronous teledata protocol is in the same state as the standard version:
\begin{equation}
\begin{aligned}
\ket{\Psi} = \frac{1}{2}(&\ket{00}_A(\alpha\ket{0} + \beta\ket{1}) + \\
			 &\ket{01}_A(\alpha\ket{1} + \beta\ket{0}) + \\
			 &\ket{10}_A(\alpha\ket{0} - \beta\ket{1}) + \\
			 &\ket{11}_A(\alpha\ket{1} - \beta\ket{0}))
\end{aligned}
\end{equation}
In Section \ref{TTSection}, a CNot operator is added between the second entangled photon and the qubit $B$ which is currently in the $\ket{0}$ state. Next, an arbitrary unitary operator is added to qubit $B$ resulting in the state
\begin{equation}
\begin{aligned}
\ket{\Psi} = \frac{1}{2}(&\ket{00}_A(\alpha \ket{0}U\ket{0} + \beta \ket{1}U\ket{1}) + \\
			 &\ket{01}_A(\alpha \ket{1}U\ket{1} + \beta \ket{0}U\ket{0}) + \\
			 &\ket{10}_A(\alpha \ket{0}U\ket{0} - \beta \ket{1}U\ket{1}) + \\
			 &\ket{11}_A(\alpha \ket{1}U\ket{1} - \beta \ket{0}U\ket{0}))
\end{aligned}
\end{equation}
To replace the role of the $X$ gate in the teledata protocol we will need an operator, which we shall denote $F$, to convert the state 
\begin{equation}
\alpha \ket{1}U\ket{1} + \beta \ket{0}U\ket{0}
\end{equation}
into the state
\begin{equation}
\alpha \ket{0}U\ket{0} + \beta \ket{1}U\ket{1}
\end{equation}
We will also need an operator to replace $Z$ in teledata. This operator, which we shall denote $G$, is tasked with converting the state
\begin{equation}
\alpha \ket{0}U\ket{0} - \beta \ket{1}U\ket{1}
\end{equation}
into the state
\begin{equation}
\alpha \ket{0}U\ket{0} + \beta \ket{1}U\ket{1}
\end{equation}

First. we will look at the operator $F$. Since the goal is to replace the $X$ gate, $F$ will need to be able to convert the state of qubit $B$ and the entangled qubit at $B$ from the state
\begin{equation} \label{eq:Fnot}
\begin{aligned}
\ket{\Psi} &= \alpha\ket{1}U\ket{1} + \beta\ket{0}U\ket{0} \\
&= \alpha\ket{1}(U^1_{\alpha}\ket{0} + U^1_{\beta}\ket{1}) + \beta\ket{0}(U^0_{\alpha}\ket{0} + U^0_{\beta}\ket{1}) \\
&= \begin{bmatrix}
\beta U^0_{\alpha}\\
\beta U^0_{\beta}\\
\alpha U^1_{\alpha}\\
\alpha U^1_{\beta}
\end{bmatrix}
\end{aligned}
\end{equation}
into the state
\begin{equation} \label{eq:F}
\begin{aligned}
\ket{\Psi} &= \alpha\ket{0}U\ket{0} + \beta\ket{1}U\ket{1} \\
&= \alpha\ket{0}(U^0_{\alpha}\ket{0} + U^0_{\beta}\ket{1}) + \beta\ket{1}(U^1_{\alpha}\ket{0} + U^1_{\beta}\ket{1}) \\
&= \begin{bmatrix}
\alpha U^0_{\alpha}\\
\alpha U^0_{\beta}\\
\beta U^1_{\alpha}\\
\beta U^1_{\beta}
\end{bmatrix}
\end{aligned}
\end{equation}
Here, $U^i$ is $U$ applied to the state $\ket{i}$, while the subscripts represent the changes of the state $\ket{i}$ towards $\ket{0}$ or $\ket{1}$ after $U^i$ was applied. For example, $U^0_{\beta}$ is the change from the state $\ket{0}$ to state $\ket{1}$ by applying $U$. Given the above states, the matrix $F$ will be
\begin{equation}
\begin{bmatrix}
\frac{\alpha}{\beta}&0&0&0\\
0&\frac{\alpha}{\beta}&0&0\\
0&0&\frac{\beta}{\alpha}&0\\
0&0&0&\frac{\beta}{\alpha}
\end{bmatrix}
\end{equation}
The first thing that needs to be pointed out is that $F$ is \emph{not} unitary. This is an area of research that has been gaining more attention in recent years \cite{zheng2021, schlimgen2022}. In particular the proposed improved teledata protocol shown in Figure \ref{fig:async_operations}, and explained below, uses a nonunitary operator $F$ which has a similar format to the nonunitary matrices proposed by Schlimgen \textit{et al}. \cite{schlimgen2022}. This matrix rescales, in a complementary way, the amplitudes of two different qubits, an action that can potentially be achieved in a physical setting by over-emphasising half of the quantum register and under-emphasising the other half, or by performing some form of measurement. More work is needed to make $F$ into a viable operator, and speculation into the physics and device engineering necessary to design this operator in a reasonable way is outside the scope of this paper. There is a single qubit version of $F$ that is also diagonal, but since it is dependant on $\alpha$, $\beta$, $U^0$, and $U^1$, it does not make as concise of a matrix due to its dependence on $U$. However, when entanglement is added between the second entangled qubit and $B$, and $F$ is made as a two qubit operator, the dependency on $U$ cancels out. \par

The operator $G$ is meant to replace the $Z$ operation in the teledata protocol --- we will follow a process that resembles the derivation of $F$. We will need to convert the state of qubit $B$ and the entangled state from
\begin{equation}
\begin{aligned}
\ket{\Psi} &= \alpha \ket{0}U\ket{0} - \beta \ket{1}U\ket{1} \\
&= \alpha\ket{0}(U^{0}_{\alpha}\ket{0} + U^{0}_{\beta}\ket{1}) + \beta\ket{1}(U^{1-}_{\alpha}\ket{0} + U^{1-}_{\beta}\ket{1}) \\
&= \begin{bmatrix}
\alpha U^{0}_{\alpha}\\
\alpha U^{0}_{\beta}\\
\beta U^{1-}_{\alpha}\\
\beta U^{1-}_{\beta}
\end{bmatrix}
\end{aligned}
\end{equation}
into the state
\begin{equation}
\begin{aligned}
\ket{\Psi} &= \alpha \ket{0}U\ket{0} + \beta \ket{1}U\ket{1} \\
&= \alpha\ket{0}(U^{0}_{\alpha}\ket{0} + U^{0}_{\beta}\ket{1}) + \beta\ket{1}(U^{1+}_{\alpha}\ket{0} + U^{1+}_{\beta}\ket{1}) \\
&= \begin{bmatrix}
\alpha U^{0}_{\alpha}\\
\alpha U^{0}_{\beta}\\
\beta U^{1+}_{\alpha}\\
\beta U^{1+}_{\beta}
\end{bmatrix}
\end{aligned}
\end{equation}
in which $U^{1+}$ and $U^{1-}$ are the results of applying $U$ to positive or negative $\ket{1}$. The resulting matrix representative for $G$ is 
\begin{equation}
\begin{bmatrix}
1&0&0&0\\
0&1&0&0\\
0&0&\frac{U^{1+}_{\alpha}}{U^{1-}_{\alpha}}&0\\
0&0&0&\frac{U^{1+}_{\beta}}{U^{1-}_{\beta}}
\end{bmatrix}
\end{equation}
The matrix $G$ retains its dependency on the operator $U$ when scaled to two qubits unlike the operator $F$. As a result, in the proposed asynchronous teledata protocol the sequence $U^{-1}ZU$ is applied to make sure that the resulting state is calculated correctly. While this is situationally less efficient than the standard teledata protocol, when the conditional $Z$ operation is not required, asynchronous operations are possible. \par 

The asynchronous telegate operation does involve some additional changes. For one, quantum device $B$ requires one additional qubit. Qubits are referenced with subscripts based on their index. Second, the telegate operation is a remote CNot operation. For it to be truly asynchronous, unitary operators need to be applied to both the states $\ket{\psi}$ and $\ket{\phi}$. In Figure \ref{fig:async_operations} we only show the synchronous version of the protocol, but the method used to apply the conditional $Z$ gate in teledata could also be applied to qubit $A$ in telegate to allow the control qubit to be semi-asynchronous. In the asynchronous telegate circuit, $B_1$ is initialized to $\ket{0}$ with $A$ and $B_0$ being in the arbitrary states $\ket{\psi}$ and $\ket{\phi}$ respectively.
\begin{equation}
\begin{aligned}
\ket{\Psi} = \frac{1}{2}(&\ket{00}_e(\alpha\ket{0}\ket{\phi} + \beta\ket{1}\bar{\ket{\phi}}) + \\
			 &\ket{01}_e(\alpha\ket{0}\ket{\phi} - \beta\ket{1}\bar{\ket{\phi}}) + \\
			 &\ket{10}_e(\alpha\ket{0}\bar{\ket{\phi}} + \beta\ket{1}\ket{\phi}) + \\
			 &\ket{11}_e(\beta\ket{1}\ket{\phi} - \alpha\ket{0}\bar{\ket{\phi}}))
\end{aligned}
\end{equation}
After applying the CNot gates and unitary operators in Section \ref{TTSection}, the state is
\begin{equation}
\begin{aligned}
\ket{\Psi}& = \frac{1}{2}( \\
			 &\ket{00}_e(\alpha \ket{0}\ket{\phi}U\ket{\phi} + \beta \ket{1}\bar{\ket{\phi}}U\bar{\ket{\phi}}) + \\
			 &\ket{01}_e(\alpha \ket{0}\ket{\phi}U\ket{\phi} - \beta \ket{1}\bar{\ket{\phi}}U\bar{\ket{\phi}}) + \\
			 &\ket{10}_e(\alpha \ket{0}\bar{\ket{\phi}}U\bar{\ket{\phi}} + \beta \ket{1}\ket{\phi}U\ket{\phi}) + \\
			 &\ket{11}_e(\beta \ket{1}\ket{\phi}U\ket{\phi} - \alpha \ket{0}\bar{\ket{\phi}}U\bar{\ket{\phi}}))
\end{aligned}
\end{equation}
The state will then need to be converted from the state
\begin{equation}
\begin{aligned}
\ket{\Psi} &= \alpha \bar{\ket{\phi}}U_B\bar{\ket{\phi}} + \beta \ket{\phi}U_B\ket{\phi}\\
&= \alpha\bar{\ket{\phi}}(U^{\bar{\phi}}_{\alpha}\ket{0} + U^{\bar{\phi}}_{\beta}\ket{1}) +
	\beta\ket{\phi}(U^{\phi}_{\alpha}\ket{0} + U^{\phi}_{\beta}\ket{1})
\end{aligned}
\end{equation}
into the state
\begin{equation}
\begin{aligned}
\ket{\Psi} &= \alpha\ket{\phi}U_B\ket{\phi} + \beta\bar{\ket{\phi}}U_B\bar{\ket{\phi}}\\
&= \alpha\ket{\phi}(U^{\phi}_{\alpha}\ket{0} + U^{\phi}_{\beta}\ket{1}) +
	\beta\bar{\ket{\phi}}(U^{\bar{\phi}}_{\alpha}\ket{0} + U^{\bar{\phi}}_{\beta}\ket{1})
\end{aligned}
\end{equation}
As for the states defined in equations \ref{eq:Fnot} and \ref{eq:F}, the operator $F$ can be used to change the phase of qubit based on $\alpha$ and $\beta$. There is one notable difference between the use of $F$ in teledata and telegate, in teledata $\alpha$ and $\beta$ are the phases in the second entangled photon which $F$ is operating on, where in the telegate $\alpha$ and $\beta$ are the phases in qubit $A$ which $F$ is not operating on. While exploring the implications of this difference in an actual implementation of $F$ is outside the scope of this paper, it is something worth mentioning.

\section{Benefits and Limitations}
In classical computing, asynchronous operations allow for multiple operations to be done in parallel hiding the cost of certain operations. In the standard teledata and telegate operations, both quantum devices involved in the process are unable to do any operations on the involved qubits while the teledata or telegate operations are taking place. With the short decoherence times of NISQ devices, having the latencies for distributing entangled photons and classical information added to the cost of an operation makes telegate and teledata very expensive. The advantage of the asynchronous teledata and telegate protocols is that the quantum devices can continue operations on the target qubits while the fourth segment of the protocols is still being executed, as was shown in Section \ref{AsyncSection}. This allows for the cost of measurement, classical communication, and the conditional operations to be hidden behind local operations. \par

There are some limitations and costs to using the asynchronous protocols. First, given the way in which telegate and teledata require the conditional $Z$ operator, the sequence $U^{-1}ZU$ needs to be applied, making the asynchronous version slower than the original in some scenarios. If the conditional $Z$ gate is not required the asynchronous versions will be faster. How often the $Z$ gate is required will be based on the initial state, so the use of the asynchronous versions of telegate and teledata will be application specific. Second, even though local operations can be performed while the asynchronous protocols are in process, any measurements on the target qubits --- or on any qubits entangled with them --- will need to be delayed until the asynchronous operations are complete. If the quantum circuit has all of its measurements at the end, the delay of classical communication will not have any noticeable effect on the outcome so long as the classical communication finishes before the qubits de-cohere. In this optimal scenario the cost of classical communication can be almost completely hidden. On the other extreme, if measurement happens right after the remote operation, then the delay of asynchronous operations will be the same as the standard telegate and teledata. In this worst-case scenario, it can more efficient to use the standard protocols from the use of $U^{-1}ZU$. Due to classical operations needing to occur before measurement, additional bookkeeping will be required to keep track of the target qubit and its entanglement to ensure that measurements do not happen out of order.

\section{Proposed System}
\begin{figure}
	\centering
	\includegraphics[scale=0.4]{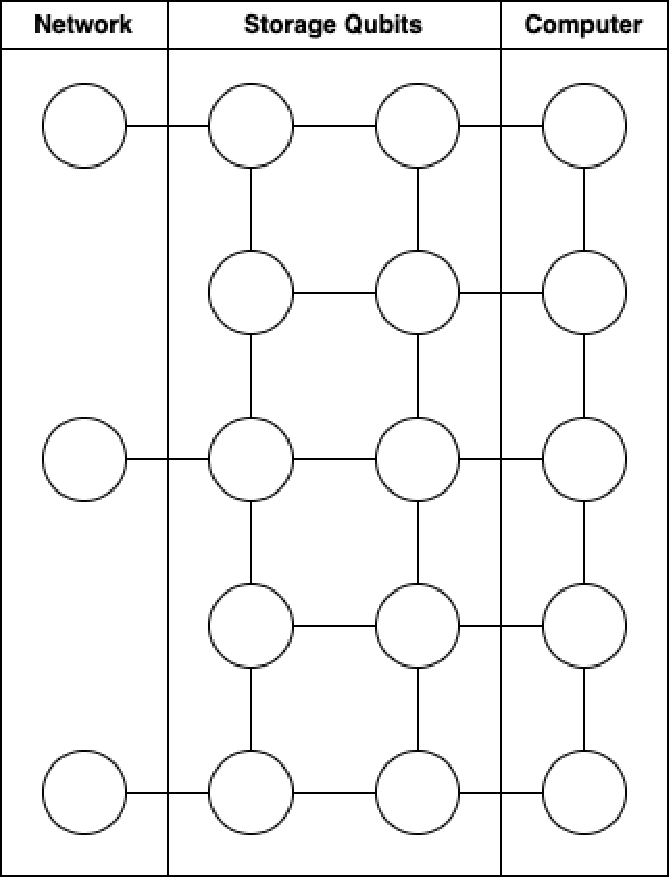}
	\caption{\textit{This is a diagram of a quantum network card. The circles represent qubits, and the lines are connection between them.}}
	\label{fig:network_card}
\end{figure}
As an example of how the asynchronous protocols could be used, a design for a basic QNC will be proposed. The QNC shown in Figure \ref{fig:network_card} contains three sections: the network section, storage qubits section, and the computer section. The network section qubits are responsible for receiving and performing operations on entangled photons transmitted over the network. The storage qubits section stores ancilla qubits used in the asynchronous protocols that need to last for the duration of the operation. The computer section represents the rest of the qubits in the quantum device, but to save space only the ones directly connected to the network card are shown. The specific number of qubits in each section and the connections are arbitrary. The main concept is having qubits dedicated to each of the three sections. \par

In order to keep the qubits in the network section of the QNC free, it is possible to use a SWAP operation to move the state of entangled photons to the qubits in the storage section. However, SWAP operations are also not cheap, but if they are used on entangled photons, that may have to wait for an arbitrary time. For example, an entangled photon in the sending device for a asynchronous teledata operation will need to be stored until the classical communication finishes. Then the cost of the SWAP operation might be worth while. The cost of SWAP operations might also be useful, or even required, for the asynchronous teledata protocol since it requires doing $U$ operations on the entangled photon at $B$. Due to variations in hardware, it is possible that the cost of doing operations on the quantum device could be noticeably cheaper than doing them directly on the entangled photon.

\section{Conclusion}
Asynchronous telegate and teledata protocols have a lot of potential to reduce the effects of classical latency on distributed quantum operations. We have described how the telegate and teledata protocols can be extended to allow for asynchronous classical communication using nonunitary operators, as well as the benefits and limitations of these protocols. With additional work, it is conceivable that efficient asynchronous distributed quantum operations could be created.

\section*{Acknowledgments}
The first- and third-named authors thank the organizers of Quantum Days 2024, held in Calgary, Canada, for creating a stimulating environment at the interface of academia and industry.  Some formative moments in the development of this work occurred during the conference.

\bibliographystyle{IEEEtran}
\bibliography{IEEEabrv,ref}

\end{document}